\documentclass[12pt]{article}

\def\bea{\begin{eqnarray}}
\def\eea{\end{eqnarray}}
\def \t{\tilde t}

\def\g {{1 \over \sqrt{1-{u^2 \over c^2 }}}}

\newcommand{\reef}[1]{(\ref{#1})}
\newcommand{\rom}[1]{\mathrm{#1}}

\usepackage{amsfonts}
\usepackage{amssymb}
\usepackage{epsfig}

\newcommand{\beqn}{\begin{eqnarray}}
\newcommand{\eeqn}{\end{eqnarray}}
\newcommand{\non}{\nonumber \\}

\newcommand{\ca}{{\cal A}}

\newcommand{\cn}{{\cal N}}

\newcommand{\cl}{{\cal L}}

\newcommand{\cg}{{\cal G}}

\newcommand{\tr}{{\rm tr}}
\newcommand{\FI}{{Fayet-Iliopoulos }}
\newcommand{\bb}[2]{\!\left[^{#1}_{#2}\right]\!}

\newcommand{\be}{\begin{equation}}
\newcommand{\ee}{\end{equation}}
\newcommand{\nn}{\nonumber}

\def\be{\begin{equation}}
\def\ee{\end{equation}}
\def\bea{\begin{eqnarray}}
\def\eea{\end{eqnarray}}
\def\ba{\begin{array}}
\def\ea{\end{array}}
\def\bd{\begin{displaymath}}
\def\ed{\end{displaymath}}

\def\nn{\nonumber}

\def\tr{{\rm tr}}

\def\unit{1 \hskip-.3em \raise2pt\hbox{$ \scriptstyle |$ } }
\def\a{\alpha}
\def\b{\beta}

\def\d{\delta}
\def\e{\epsilon}           % Also, \varepsilon
\def\g{\gamma}

\def\l{\lambda}
\def\m{\mu}
\def\n{\nu}
  
\def\th{\theta}                  %     \vartheta
\def\r{\rho}                                     %     \varrho
\def\s{\sigma}                                   %     \varsigma
\def\t{\tau}

\def\x{\xi}

\def\G{\Gamma}

\def\L{\Lambda}
\def\O{\Omega}

\def\S{\Sigma}

\def\bop#1{\setbox0=\hbox{$#1M$}\mkern1.5mu
        \vbox{\hrule height0pt depth.04\ht0
        \hbox{\vrule width.04\ht0 height.9\ht0 \kern.9\ht0
        \vrule width.04\ht0}\hrule height.04\ht0}\mkern1.5mu}
\def\pa{\partial}                              % curly d
\def\>{\rangle} %right angle
\def\<{\langle} %left angle
\def\Dsl{D \hskip-.6em \raise1pt\hbox{$ / $ } }
\def\to{\rightarrow}

\def\pa{\partial}

\def\+{\oplus}

\def\lab{\label}
\def\sq2{\sqrt{2}}

\def\ba{{\bar{\a}}}
\def\bb{{\bar{\b}}}

\def\bd{\bar{\d}}

\begin{document}
%%%%%%%%%%%%%%%%%%%%%%%%%%%%%%%%%%%%%%%%%%%%%

\thispagestyle{empty}

%% HE May 25, 2006

~
\begin{flushright}
\vspace{-3cm}
{\small 
MIT-CTP-3743\\
CERN-PH-TH/2006-086
}
\end{flushright}

\vspace{1.5cm}

\begin{center}
{\bf\LARGE Anomaly Cancellation in Supergravity \\[.3cm] with \FI Couplings}

\vspace{1.5cm}

{\bf Henriette Elvang$^1$,\, Daniel Z.~Freedman$^{1,2}$\,
and\, Boris K\"ors}$^3$ \vspace{1cm}

{\it
$^1$Center for Theoretical Physics, Laboratory for Nuclear Science \\
and Department of Physics, Massachusetts Institute of Technology \\
Cambridge, Massachusetts 02139, USA \\

$^2$Department of Mathematics, Massachusetts Institute of Technology \\
Cambridge, Massachusetts 02139, USA \\

$^3$Physics Department, Theory Division, CERN\\ 
CH-1211 Geneva 23, Switzerland 
}

\vspace{1cm}

{\bf Abstract}
\end{center}\vspace{-.3cm}
We review and clarify the cancellation conditions for gauge anomalies
which occur when $\cn=1,~D=4$ supergravity is coupled to a K\"ahler non-linear
sigma-model with gauged isometries and \FI couplings. 
For a flat sigma-model target space and vanishing \FI couplings, 
consistency requires just the conventional anomaly cancellation
conditions.
A consistent model with
non-vanishing \FI couplings is unlikely unless the
Green-Schwarz mechanism is used.
In this case the $U(1)$ gauge boson becomes massive and the D-term
potential receives corrections.  
A Green-Schwarz mechanism can remove both the abelian and certain
non-abelian anomalies in models with a gauge non-invariant K\"ahler
potential.

\vspace{-.5cm}

\clearpage
\tableofcontents

%%%%%%%%%%%%%%%%%%%%%%%%%%%%%%%%%%%%%%%%%%%%%%%%%%%
\setcounter{equation}{0}
\section{Introduction}

In $d=4$, $\cn=1$ supergravity fermions couple to 
the K\"ahler connection and the sigma-model
connection. These are composite connections constructed from
elementary scalars and gauge potentials.
Since the gauge transformations are
embedded in the geometry of a sigma-model manifold these
connections are in general not gauge invariant.
In particular the K\"ahler connection acts as an
additional abelian gauge field that effectively gauges the $U(1)_R$
symmetry. Fermions have chiral couplings to both elementary and composite
connections, so there may be anomalies which threaten the consistency
of the theory. This subject has been investigated both in supergravity
models, see \cite{Freedman:1976uk,drei,cast} for early work, and
string compactifications
\cite{LopesCardoso:1991zt,Derendinger:1991hq,Kaplunovsky:1994fg}.

A general analysis 
of the quantum consistency
conditions for supergravity 
was recently presented in 
\cite{Freedman:2005up}, in which effects of the composite
connections were emphasized. 
It follows from the gauge field equations of motion
$D_\mu {F^a}^{\mu\nu} = {J^a}^\nu$ that
\be
 \label{genConsist}
  0 \equiv D_\nu D_\mu {F^a}^{\mu\nu} = D_\nu {J^a}^\nu  \, .
\ee
The left side vanishes identically, so the current ${J^a}^\nu$ must be
conserved. Conservation holds in the classical theory, but can be
violated 
in the quantum theory by anomalies, 
viz.~$D_\nu {J^a}^\nu = \mathcal{A}^a \ne 0$.
The quantum theory is inconsistent unless the anomalies are cancelled.
The detailed consistency conditions for supergravity were found in 
\cite{Freedman:2005up} and expressed in terms of covariant anomalies.  

The purpose of this paper is to unravel the structure of the
consistency conditions and clarify the anomaly cancellation conditions
required by quantum consistency of the theory. 
We focus here on theories with flat target spaces. We recast the
consistency conditions in terms of consistent anomalies and consider
the effects of finite local counter terms. 
This reduces the results of \cite{Freedman:2005up} to a set of
physically necessary consistency conditions. We further discuss the
Green-Schwarz mechanism, which requires additional degrees of freedom.

Two distinct consistency conditions, abelian and non-abelian, arise
from \reef{genConsist} depending on whether the current ${J^a}^\nu$ is
abelian or non-abelian. 
We are especially concerned with \emph{mixed anomalies}. 
Using $C_{\m\n}$ and $F^a_{\m\n}$ for the abelian and non-abelian
field strengths,  
examples of (covariant) mixed anomalies are terms involving
the non-abelian fields like $\epsilon^{\m\n\r\s} \tr\,F_{\m\n}
F_{\r\s}$ in the  
divergence of the abelian current (``mixed abelian anomalies''), or 
$\epsilon^{\m\n\r\s} \tr\, T^a C_{\m\n} F_{\r\s}$ in the 
divergence of the non-abelian current (``mixed non-abelian
anomalies'').
The field strength $K_{\mu\nu} = \pa_\mu K_\nu - \pa_\nu K_\mu$ of the
K\"ahler connection $K_\mu$ also appears in the mixed anomalies.

For supergravity theories with a flat sigma-model target space and a
gauge group $G\times U(1)$ we  
clarify the anomaly cancellation conditions, with and
without \FI couplings.
The results are 
summarized and discussed at the end of section \ref{Kan}. In case of
vanishing \FI couplings, standard conditions on the matter content
suffice to ensure consistency of the theory.
But for general non-vanishing \FI couplings
consistency requires the Green-Schwarz mechanism. 

The standard Green-Schwarz mechanism can only
remove anomalies in abelian conservation laws.\footnote{An exception
  to this is studied in section \ref{noninvK}.}  
Consequently, to ensure consistency, all mixed non-abelian
anomalies must be cancelled by other means; either by finite local
counter terms or by imposing conditions on the matter content of the
theory. 
Furthermore, the Green-Schwarz mechanism requires that the abelian
anomaly removed is gauge covariant.
However, not all consistent mixed abelian anomalies 
are covariant. 
Local counter terms are needed to restructure them
before the Green-Schwarz method is applied.  
We construct here finite local counter terms which have
both required properties: (1) they remove completely the
consistent mixed non-abelian anomalies, and (2) they simultaneously
convert the consistent mixed abelian anomalies to covariant form.
The final resulting consistency conditions are summarized in section
\ref{s:GS}, where we also comment on the effect of gravitational
anomalies. 

We focus in most of this work on K\"ahler potentials that are
invariant under non-abelian gauge transformations. A simple model with
a non-invariant K\"ahler potential is studied in section
\ref{noninvK}. Since the K\"ahler connection transform as a
$U(1)$ connection under non-abelian gauge transformations, 
a Green-Schwarz mechanism can be used to
cancel certain anomalies in the non-abelian current
conservation law.

This study is relevant to various models of cosmology and particle 
physics that make use of \FI couplings, such as D-term
inflation \cite{Binetruy:1996xj,Halyo:1996pp}, or supersymmetry
breaking via an anomalous $U(1)$ \cite{Dvali:1996rj}, or the
string solutions for so-called 
D-strings in
\cite{Binetruy:1998mn,Dvali:2003zh,Davis:2005jf,Blanco-Pillado:2005xx},
see also \cite{Binetruy:2004hh,Lawrence:2004sm}. 
In string theory 
\FI couplings were discussed by Dine, Seiberg and Witten
\cite{Dine:1987xk}.  
More recently, D-terms and \FI couplings have played an important role in 
the context of moduli stabilization in string compactifications, see
\cite{vz}, 
following the proposal of Kachru, Kallosh, Linde, and Trivedi
\cite{Kachru:2003aw}. Our 
conclusion about the validity of such models at the quantum level
is that \FI couplings can only be consistent if one incorporates a
Green-Schwarz mechanism to cancel residual anomalies. Such a
Green-Schwarz mechanism modifies the physics by generating a mass for
the gauge boson and by a contribution to the D-term potential. 

The paper is organized as follows. In section \ref{s:anomalies}
we review the covariant and consistent anomalies, and we show how
finite local counter terms can be used to restructure mixed anomalies.
We review in section \ref{Kan} the consistency conditions derived in
ref.~\cite{Freedman:2005up}, write the anomaly cancellation
conditions in terms of the consistent anomaly, and include
local counter terms to reduce the anomaly cancellation conditions.
The Green-Schwarz mechanism is
discussed in section \ref{s:GS}. 
We discuss briefly in section \ref{s:more} generalizations to
non-gauge invariant K\"ahler potentials as well as the
supersymmetrization of anomalies and local counter terms. 
We conclude in section \ref{s:concl} with a
summary and discussion of our results.

%%%%%%%%%%%%%%%%%%%%%%%%%%%%%%%%%%%%%%%%%%%%%%%%%%%

\setcounter{equation}{0}
\section{Covariant vs.~consistent anomalies}
\label{s:anomalies}

To derive the physically relevant anomaly cancellation conditions from
the requirement of current conservation it is crucial to use the
proper form of the anomalies and include all possible (finite) local
counter terms into the Lagrangian. To provide the background material
needed later we have included a short review section on consistent and
covariant anomalies.

Consider the kinetic Lagrangian of a Weyl fermion, written as the
left-handed component of a Dirac fermion, minimally coupled to a
background gauge field $V_\m$,\footnote{We concentrate on the
left-chiral part of the gauge transformations. Generalizations are
straight forward.} 
\be 
  \label{genL}
   \cl 
   \;= \; \bar{\psi} \gamma^\mu D_\mu L\psi
\ , \quad 
D_\mu  L\psi ~=~ (\pa_\mu + V_\mu) L \psi \ , 
\ee
where $L=\frac{1}{2}(1 - \gamma_5),\ R=\frac{1}{2}(1+\gamma_5)$ and
$V_\mu = V_\mu^a T^a$. The $T^a$ are anti-hermitian generators of a
Lie algebra for a gauge group $\cg$, which may contain $U(1)$
factors. The action with \reef{genL} is invariant under chiral gauge
transformations, 
\beqn
  L \psi &\to& e^{-\theta(x)}L \psi \, , \qquad
  V_\mu \to  e^{-\theta(x)}V_\mu e^{\theta(x)}
  +  e^{-\theta(x)}\pa_\mu e^{\theta(x)}
\eeqn 
with $\theta=\theta^a T^a$. Correspondingly, the left-chiral current  
is classically conserved,
\beqn\lab{Lcur}
j^{a}_\mu = - \bar{\psi} \gamma_\mu T^a L \psi\, ,  \quad 
0=D_\mu j^{\mu}=\pa_\mu j^{\mu} + [V_\mu,j^{\mu}] \, .
\eeqn
Classical symmetries and conservation laws
receive corrections in the quantum theory due to anomalies.

%%%%%%%%%%%%%%%%%%%%%%%%%%%%%%%%%%%%%%%%%%%

\subsection{Anomalies}

The covariant (left-)chiral anomaly is
\bea
  \nonumber
  (D_\mu j^{\mu})^a &=& \frac{i}{32\pi^2} \e^{\m\n\r\s} 
   \tr\Big[ T^a V_{\m\n}V_{\r\s} \Big] \\
  &=& \frac{i}{8\pi^2} \e^{\m\n\r\s} 
      \tr\Big[ T^a \pa_\mu
      \Big( 
        V_\nu \pa_\r V_\s + \frac{2}{3} V_\n V_\r V_\s
      \Big) \Big]
      \, , 
  \label{cov1}
\eea
with $V_{\m\n} = \pa_\m V_\n - \pa_\n V_\m + [V_\m,V_\n]$.  
An anomaly reflects the gauge non-invariance of the effective action 
\be
   e^{-W[V_\mu]} = \G[V_\m] 
  = \int \mathcal{D}\bar{\psi}\mathcal{D}\psi \, 
  e^{-S[V_\m,\bar{\psi},\psi]} \, ,
\ee
where $V_\m$ denotes a background gauge field. Defining the current 
$j^{a\,\mu}=- \d \cl [V_\m]/\d V^a_\mu$ one has 
\be
  \d_\theta W[V_\m] 
  = \int d^4x\ \theta^a \mathcal{A}^a \, ,
\ee
where $\mathcal{A}^a= \langle (D_\mu j^{\mu})^a \rangle$ is the anomaly under the gauge transformation
\beqn \lab{gvar}
\d_\th V_\m = D_\m \th = \pa_\m\th + [V_\m,\th] \ .
\eeqn 
The Wess-Zumino consistency condition 
$[\d_{\theta_1},\d_{\theta_2}]W[V_\m]  =
\d_{[\theta_1,\theta_2]}W[V_\m]$ requires 
\be
  \label{WZcond}
  \d_{\theta_1} (\theta_2^a \mathcal{A}^a) 
  - \d_{\theta_2} (\theta_1^a \mathcal{A}^a) 
  = [\theta_1,\theta_2]^a \mathcal{A}^a \,  
\ee
for the anomaly \cite{Wess:1971yu}. It is not satisfied for a simple
non-abelian gauge 
group by the covariant anomaly \reef{cov1}, since a factor of 2 appears
on the right side of eq.~\reef{WZcond}, see e.g.\
\cite{Bertlmann:1996xk}.

The form of the anomaly that does satisfy \reef{WZcond} is known as
the \emph{consistent anomaly} (Bardeen \cite{Bardeen:1969md}, Gross
and Jackiw \cite{Gross:1972pv}). 
The consistent anomaly follows from a Bose symmetric regularization of
the triangle and quadrangle Feynman diagrams for correlation functions
of the potentials $V_\m$.
For the left-chiral current \reef{Lcur} it
can be written\footnote{For a right-chiral current the overall sign of
the anomaly changes.}
\beqn
  \label{GenCon1}
  (D_\mu j^{\mu})^a 
  = \frac{i}{24\pi^2} \e^{\m\n\r\s} \tr\Big[ T^a\, \pa_\mu
  \Big(
    V_\n\, \pa_\r V_\s 
    + \frac{1}{2} V_\n V_\r V_\s 
  \Big)\Big] \, .
\eeqn
Note that \reef{cov1} does not transform covariantly.
For abelian $V_\mu$, the cubic term is absent and the anomalies
\reef{cov1} and \reef{GenCon1} 
differ by an overall factor of $\frac13$
which accounts for Bose symmetry of the triangle amplitude. 

The reason Wess-Zumino consistency fails for the covariant anomaly is
that the current on the left-hand side of \reef{cov1} is not the
variation of the effective action. Bardeen and Zumino showed
\cite{Bardeen:1984pm} that the current in \reef{cov1} differs from
the consistent current by a
polynomial local in the gauge potential. For a simple non-abelian gauge 
group the Bardeen-Zumino polynomial cannot be written as the gauge
variation of a finite local 
counter term added to the effective action. Hence the covariant and
the consistent anomalies are not physically
equivalent.\footnote{Nonetheless, 
 the vanishing of the covariant and the consistent anomalies
 for a simple gauge group requires the same condition, namely
  $\tr\,T^a\{T^b,T^c\}=0$.} 
For the purpose of analyzing the consistency conditions
\cite{Freedman:2005up}   
for currents which are sources of gauge fields, the relevant form of
the anomaly is the consistent anomaly \reef{GenCon1}.

%%%%%%%%%%%%%%%%%%%%%%%%%%%%%%%%%%%%%%%%%%%%%%%%%%%%%%%%%%%%%%%%%%%%%%%

\subsection{Local counter terms}
\label{s:lct}

Having established that the two forms, \reef{cov1} and \reef{GenCon1},
of anomalies are not equivalent 
for a simple non-abelian gauge group, we point out how one can
interpolate between them for a mixed anomaly.\footnote{Recent
  interesting work \cite{Anastasopoulos:2006cz} studies local counter
  terms and the Green-Schwarz mechanism in connection to anomalous
  $U(1)$'s.}
This will later be
crucial for applying the Green-Schwarz mechanism.

Consider a gauge group which is the product of a single $U(1)$
factor and a simple non-abelian group $G$, $\cg=G\times U(1)$. Write
the gauge field $V_\mu = A_\mu^a T^a + i Q C_\mu$, where the $T^a$ are
anti-hermitian generators of $G$ and $Q$ is the charge under $U(1)$. 
We use $F^a_{\m\n}$ for the non-abelian field strength and $C_{\m\n}$
for the abelian field strength. 
Inserting $V_\mu = A_\mu^a T^a + i Q C_\mu$ into the expression for 
the consistent anomaly \reef{GenCon1} and covariant anomaly
\reef{cov1}, we pick up terms which are purely 
abelian or purely non-abelian anomalies as well as mixed anomalies.
We write this 
\beqn 
(D_\m j^\m)^a &=& \ca^a ~=~ \ca^a_{\rm non-ab} + \ca^a_{\rm mixed} \ , 
\non
(D_\m j^\m)^Q &=& \ca^Q ~=~ \ca^Q_{\rm abel} + \ca^Q_{\rm mixed} \ , 
\eeqn 
where $j_\m^a = -\d \cl/\d A^a_\m= -i\bar{\psi} \gamma_\mu T^a
L\psi$ and $j_\m^Q = -\d \cl/\d C_\m = -i\bar{\psi} \gamma_\mu iQ\,
L\psi$ are the non-abelian and abelian currents.  
Below subscripts ``cov'' or ``con'' indicate whether a given term in the
anomalies is written in the covariant form \reef{cov1} or the
consistent form \reef{GenCon1}. 

To be explicit, we list the mixed anomalies $\ca^a_{\rm mixed}$
and $\ca^Q_{\rm mixed}$ in covariant and consistent form respectively,
\beqn
  \nn
  \mathcal{A}^Q_\rom{mixed\;cov}
  &=&  \frac{i}{32\pi^2} \e^{\m\n\r\s}\tr\,iQ\,F_{\m\n}F_{\r\s} =  
  \frac{i}{8\pi^2} \e^{\m\n\r\s}\tr\Big[ iQ \pa_\mu
  \Big(
    A_\n\,  \pa_\r A_\s 
    + \frac{2}{3} A_\n A_\r A_\s
  \Big) \Big] \, ,
\\[2mm] \nonumber
  \mathcal{A}^Q_\rom{mixed\;con}
  &=&  
  \frac{i}{24\pi^2} \e^{\m\n\r\s}\tr\Big[ iQ\pa_\mu
  \Big(
    A_\n\,  \pa_\r A_\s 
    + \frac{1}{2} A_\n A_\r A_\s
  \Big) \Big] \, ,
\\[2mm] \nonumber
\mathcal{A}^a_\rom{mixed\;cov}
  &=&   \frac{i}{16\pi^2} \e^{\m\n\r\s}\tr\Big[ T^aiQ
\,C_{\m\n}F_{\r\s}\Big] 
\, , 
\\[2mm]
\mathcal{A}^a_\rom{mixed\;con}
&=&  
  \frac{i}{12\pi^2} \e^{\m\n\r\s}\tr\Big[ T^aiQ\pa_\mu
  \Big(
    C_\n\,  \pa_\r A_\s 
    + \frac{1}{4} C_\n A_\r A_\s
  \Big) \Big] \, .
  \label{mixedAnom}
\eeqn

There are two candidate polynomials in $C_\mu$ and $A_\mu$
from which finite local counter terms in the Lagrangian can be constructed,   
\beqn\lab{ct}
\cl_1 &=& -\frac{i}{12\pi^2} \,
   \e^{\mu\nu\rho\sigma} C_\mu \tr\Big[ iQA_\nu\pa_ \rho A_\sigma\Big] \, , \non
  \cl_2 &=& -\frac{i}{12\pi^2} \,
   \e^{\mu\nu\rho\sigma} C_\mu \tr\Big[iQ A_\nu A_ \rho A_\sigma\Big] \, .
\eeqn
Their gauge variations under non-abelian gauge transformations
\reef{gvar} are (up to total derivatives) 
\beqn\lab{ctvar}
  \delta_\theta \cl_1 &=& -\frac{i}{12\pi^2}
  \, \e^{\mu\nu\rho\sigma}\tr\Big[ iQ\theta 
  \Big(
    \pa_\mu C_\nu\, \pa_\rho A_\sigma
    -2\pa_\mu (C_\nu A_\rho A_\sigma)
  \Big) \Big] \, , 
\non
  \delta_\theta \cl_2 &=& 
  -\frac{i}{4\pi^2}\, \e^{\mu\nu\rho\sigma}\tr\Big[ iQ\th \pa_\mu (C_\nu A_\rho A_\sigma) \Big]\, .
 \eeqn
One may add these counter terms with arbitrary coefficients
to the Lagrangian. This would modify the non-abelian current
conservation law by terms proportional to \reef{ctvar}.
The unique combination 
\beqn \label{genSct}
\cl_{\rm ct} ~=~ \cl_1+ \frac34\cl_2 
\eeqn
precisely cancels the
non-abelian mixed anomaly in the consistent form, 
\beqn \label{nonabelvarLct}
\delta_\theta \cl_{\rm ct} ~=~ -\theta^a
\mathcal{A}_\rom{mixed~con}^a\ .
\eeqn  
Under abelian gauge variations 
\beqn \lab{abvar}
\delta_\Lambda C_\mu = \pa_\mu \Lambda
\eeqn 
the counter term gives
\be
  \label{abelvarLct}
  \d_\L \cl_{\rm ct} =
-\L \big( \mathcal{A}^Q_\rom{mixed~con} 
- \mathcal{A}^Q_\rom{mixed~cov} \big) \, ,
\ee
i.e.~it ``rotates''  
the consistent form of the abelian mixed anomaly into covariant
form. As discussed in the Introduction, this is essential for the
Green-Schwarz mechanism. 

The gauge variations \reef{ctvar} of the counter terms yield total
derivatives, but the covariant mixed anomaly
$\mathcal{A}^a_\rom{mixed~cov}$ given in 
\reef{mixedAnom} involves $\e^{\m\n\r\s} \tr\, T^a C_{\mu \nu}
F_{\r\s}$ which is 
not a total derivative. Hence the non-abelian variations of the
counter terms \reef{ctvar} could never fully cancel
$\mathcal{A}^a_\rom{mixed~cov}$, 
it is therefore crucial to use $\mathcal{A}^a_\rom{mixed~con}$. 

%%%%%%%%%%%%%%%%%%%%%%%%%%%%%%%%%%
\setcounter{equation}{0}
\section{K\"ahler anomalies in supergravity}
\lab{Kan}

We start with a summary of some relevant structures of
the supergravity Lagrangian and of results of
\cite{Freedman:2005up}.

%%%%%%%%%%%%%%%%%%%%%%%%%%%%%%%%%%%

\subsection{Supergravity and composite connections} 

We consider theories with a supergravity multiplet
$(e^i_\mu,\Psi_\mu)$ coupled to gauge multiplets $(V_\mu^a,\lambda^a)$
and chiral multiplets $(z^\a,L\psi^\a)$. 
We write the gravitino $\Psi_\m$ and the gauginos $\l^a$
as four-component Majorana spinors, and
the Weyl spinors of the chiral multiplets are written with
projectors $L,R$. 
We write the action $S = \int d^4x\,\sqrt{-g} \,\cl$ and 
use $\epsilon^{0123}=(-g)^{-1/2}$. 
More details, including the Lagrangian, are given in
\cite{Freedman:2005up}.

The scalar fields are complex coordinates on a K\"ahler manifold
with K\"ahler potential $K=K(z,\bar{z})$ and metric $G_{\a\bb} =
K_{,\a\bb}$ (a comma indicating a partial, a semi-colon a covariant
derivative). 
In supergravity, isometries of the K\"ahler manifold generated by
holomorphic Killing vectors, $X^{a\alpha}(z)$, $X^{a\ba}(\bar{z})$, can be
gauged. Holomorphic Killing vectors can be expressed as gradients of a
real Killing prepotential $D^a(z,\bar{z})$ as $D^a_{\, ,\bb}=-iX^{a\alpha}
G_{\a\bb}$. 
For non-abelian gauge groups the prepotentials are uniquely determined
by the requirement that they transform in the adjoint
representation. For abelian gauge groups, there is an additional
freedom of adding a constant, $D^a \to D^a + \xi^a$. These constants
are the \FI couplings of the theory.

The K\"ahler metric must be invariant under the isometry, and this
requirement is exactly the Killing equation
\beqn
  \delta^a G_{\a\bb} = X^a_{\a;\bb} + X^a_{\bb;\a} = 0 \, .
\eeqn
However, the K\"ahler potential need not be invariant and transforms
as 
\beqn \lab{Kgauge}
\d^a K(z,\bar z) = X^{a\a} K,_\a + X^{a\bar\a} K,_{\bar\a} = F^a(z) +
\bar F^a(\bar z)\ .
\eeqn
The holomorphic function $F^a(z)$ is related to $D^a$,
\beqn \lab{Fdef}
  F^a = X^{a\a} K,_\a + i D^a \, . 
\eeqn 

The supergravity model contains an elementary gauge field $V^a_\mu$ for
each isometry. They appear in covariant derivatives of the scalars
as $D_\mu z^\a = \pa_\mu z^\a - V^a_\mu X^{a\a}$ and in the composite
K\"ahler connection
\beqn\lab{Kcon}
K_\m = \frac1{2i} \Big( K,_\a D_\m z^\a + F^a V^a_\m - \
{\rm c.c.}\ \Big)
     = \frac1{2i} \Big( K,_\a \pa_\m z^\a  - \ {\rm c.c.}\ \Big) 
     + V^a_\m D^a \ .
\eeqn
Under gauge transformations \reef{gvar} for $V^a_\mu$ and
$\delta z^\a = \theta^a(x) X^{a\a}$ for $z^\a$,
the K\"ahler connection transforms as a
$U(1)$ connection
\beqn \lab{Ktraf}
 \d K_\m = \pa_\m ( \theta^a {\rm Im}\, F^a(z)) \ . 
\eeqn 
The non-invariance of $K$ and the coupling of $K_\m$ in the covariant
derivatives of all fermions are the essential complicating factors of
the anomaly analysis in supergravity.

The fermion covariant derivatives are
\beqn \lab{dcov}
D_\mu \Psi_\nu &=& \Big( \nabla_\mu + \frac12 iK_\mu \gamma_5
\Big) \Psi_\nu ~=~ \Big( \pa_\mu + \frac14 \omega_{\mu
ij}\gamma^{ij} + \frac12 iK_\mu \gamma_5 \Big) \Psi_\nu
\ , \non
D_\mu \l^a &=& \Big( \nabla_\mu \d^{ac} + \frac12 iK_\mu
\gamma_5\d^{ac} + f^{abc} V^b_\m \Big) \l^c  \ ,
\non
D_\mu L\psi^\a &=& \Big( \nabla_\mu \d^\a_\b + \S_{\b\m}^\a
 + \frac12 iK_\mu  \d^\a_\b - X^{a\a},_\b V^a_\m \Big) L\psi^\b  \ .
\eeqn
The first line of (\ref{dcov}) defines the derivative $\nabla_\m$
which includes the spin-connection. 
The other composite connection is the sigma-model connection
$\S^\a_{\m\b} = \G^\a_{\b\g} D_\m z^\g$,
where $\G^\a_{\b\g} = G^{\a\bd}G_{\b\bd,\g}$ are the 
K\"ahler Christoffel connections. It will not be important for us. 
Note that $\frac{1}{2}K_\mu$ gauges a $U(1)_R$ symmetry under which
$L\Psi_\m$ and $L\l^a$ have charge $+1$ and $L\psi^\alpha$ has charge
$-1$.
The gravitational coupling $\kappa$ has been set to $\kappa=1$ for
simplicity, but it actually appears in the fermion covariant
derivatives through $\kappa^2 K_\mu$. 

The infinitesimal gauge transformations of the fermions are 
\beqn 
\d \Psi_\m &=& -\frac i2 \th^a {\rm Im}\, F^a(z) \g_5 \Psi_\m \ , \non
\lab{gauferm}
\d \l^a &=& f^{abc} \l^b \th^c -\frac i2 \th^b {\rm Im}\, F^b(z) \g_5
\l^a \ , \\
\nonumber
\d L\psi^\a &=& \th^a X^{a\a},_\b L \psi^\b -\frac i2 \th^a {\rm Im}\,
F^a(z) L\psi^\a \ .  
\eeqn 
The $\rom{Im}\, F^a$-terms compensate the transformation \reef{Ktraf} of
$K_\m$, while the other terms are standard transformations of gauginos
and chiral fermions. 

The chiral transformations \reef{gauferm} are anomalous and it is
these anomalies 
which are studied in \cite{Freedman:2005up}. Consistency of the
quantum theory requires that the following combination of anomalous
current divergences must cancel:
\beqn \lab{con}
0 &=& 
iY^a_{\a\bb} \< \nabla_\m (\bar\psi^\bb \g^\m L\psi^\a) \>
+ \frac12 \< \nabla_\m ( \bar\l^b f^{abc} \g^\m \l^c) \>
+ \frac12 {\rm Im}\, F^a \< \nabla_\m N^\m \>  \ ,
\eeqn
with 
\beqn 
Y^a_{\a\bb} = \frac{1}{2i} \Big( G_{\g\bar\b} X^{a\g},_\a - G_{\a\bar\g}
X^{a\bar\g},_{\bar\b} \Big)\ . 
\eeqn 
The current $N^\m$ is the $U(1)_R$ current to which the K\"ahler
connection couples, namely
\beqn\lab{noether}
N^\m = -\frac i2 \Big[ 2 G_{\a\bb} \bar\psi^\bb \g^\m L\psi^\a  + \bar\l^a \g^\m \g_5 \l^a
 + \bar\Psi_\r \g^{\r\m\n} \g_5 \Psi_\n \Big] \ .
\eeqn
In (\ref{con}) the brackets $\<...\>$ indicate 
the quantum anomalies of each current. 
These anomalies were computed in \cite{Freedman:2005up} as covariant
anomalies using the Fujikawa method.
The expressions for the
anomalies are rather 
complicated in the general case, involving the field strengths of the
full connections in (\ref{dcov}). The consistency conditions
\reef{con} will be rewritten in terms of consistent anomalies in
section \ref{s:cons}.

\subsection{Supergravity models with flat target space}

In most of this paper we will restrict the treatment to models with
flat target space and linearly realized gauge symmetries,
\beqn \lab{flat1}
K(z,\bar z) = \delta_{\a\bb} z^\a z^\bb \ , \quad
G_{\a\bb} = \d_{\a\bb}\ , \quad
X^{a\a},_\b = -T^{a\a}{}_\b = -T^{a i}{}_j e_i^\a e^j_\b \ . 
\eeqn
The sigma-model connection then vanishes, 
$\S_{\b\m}^\a = \G^\a_{\b\g} D_\m z^\g=0$. 
Although the K\"ahler potential is gauge invariant, $\rom{Im}F^a$ can
be nonzero when there are \FI couplings, i.e.\ 
\beqn
\label{flat2}
F^a = i \xi^a
\eeqn
for
abelian factors of the gauge group.
The K\"ahler connection (\ref{Kcon}) then becomes 
\beqn \lab{Kcon2}
K_\m = {\rm Im}( \d_{\a\bb} z^\bb D_\m z^\a ) + \xi^a V^a_\m\ . 
\eeqn 
The $T_{ij}^a$ are the
anti-hermitian constant matrix generators of the gauge symmetry
and we use
\beqn 
\label{LieAlg}
[T^a,T^b] = f^{abc} T^c\ , \quad  f^{acd}f^{bcd} =
\d^{ab} C_2(G)\ , \quad
{\rm tr}_r (T^aT^b) = - C(r) \d^{ab}\ ,
\eeqn
where ``tr$_r$'' indicates the trace over the irreducible
representation $r$ and ``tr'' the trace over the full spectrum of
chiral fermions, not including the gauginos or the gravitino. In this limit
(\ref{con}) simplifies to
\beqn \lab{con2}
0 ~=~ -\< \nabla_\m (\bar\psi^i T^a_{ij} \g^\m L\psi^j )\>
+ \frac12 \<\nabla_\m ( \bar\l^b f^{abc} \g^\m \l^c) \>
+ \frac12 \xi^a \< \nabla_\m N^\m \>
\ .
\eeqn
This is the consistency condition that we now study in detail. 

%%%%%%%%%%%%%%%%%%%%%%%%%%%%%%%%%%%%%%%%%%%%%%%%%%%%%%%%%%%%%%

\subsection{Consistent anomalies}
\label{s:cons}

As we have pointed out, it is the consistent form of anomalies that
are relevant to conservation laws of gauge currents.
We now
evaluate the consistency condition \reef{con2} using \reef{GenCon1}
for the anomalies. We specialize to the gauge group $G\times U(1)$. As 
in section \ref{s:lct} we use $T^a$ and $iQ$ for the generators and
$A^a_\m$ and $C_\m$ for the gauge fields, $F^a_{\m\n}$ and $C_{\m\n}$
for their field strengths. If necessary, we label abelian quantities
by $Q$, but we drop the label $a$ on the \FI coupling
$\xi$ of the single $U(1)$. This notation should not be confused with
the previous section where $a$ was an index of the full gauge
group, not just $G$. In particular, we now write $F^a=0$ and
$F^Q=i\xi$ instead of \reef{flat2}. 
With these simplifications, the gauge potentials coupling in the
left-chiral covariant derivatives \reef{dcov} are
\beqn
\Psi_\m : &&
V_\m = -\frac i2 K_\m\ ,\non
\l^a : && V^{ab}_\m = -A^c_\m f^{abc} - \frac i2 K_\m \d^{ab}\
,\non L\psi^\a : && V_\m = A^a_\m T^a + iQ C_\m +  \frac i2 K_\m\
.
\eeqn

The consistent anomalies are obtained by inserting the relevant
connection $V_\mu$ for each of the three types of fermions in
\reef{GenCon1} and collect results.

\subsubsection*{Non-abelian consistency condition:}
The non-abelian consistent anomaly of the chiral fermion current now reads
\beqn \lab{conan2a}
-\< \nabla_\m (\bar\psi^i T^{a}_{ij} \g^\m L\psi^j )\> &=&
\frac{1}{24\pi^2} \e^{\m\n\r\s} \tr\, T^a \Bigg[ 
i\Big\{ \pa_\m A_\n \pa_\r A_\s + \frac12 \pa_\m ( A_\n A_\r A_\s ) \Big\} \\
&&\hspace{-3cm} \nn
 - \Big\{ \pa_\m K_\n \pa_\r A_\s + \frac14  \pa_\m ( K_\n A_\r A_\s )
\Big\} 
 - 2 Q \Big\{ \pa_\m C_\n \pa_\r A_\s
+ \frac14 \pa_\m ( C_\n A_\r A_\s ) \Big\}
\Bigg] \ .
\eeqn
We recognize in the first line the standard purely non-abelian
anomaly and in the two other lines the mixed $G^2-U(1)$
plus the mixed $G^2-$K\"ahler anomalies. We also have 
\beqn
\lab{conan2b}\hspace{-.5cm}
\frac 12 \<\nabla_\m ( \bar\l^b f^{abc} \g^\m \l^c) \> &=&
-\frac{1}{24\pi^2} \e^{\m\n\r\s} C_2(G) 
\Big[\pa_\m K_\n  \pa_\r A_\s^a 
 + \frac18 f^{abc} \pa_\m ( K_\n A_\r^b A_\s^c )
\Big]\  .
\eeqn
\subsubsection*{Abelian consistency condition:}
The abelian consistent anomaly of the chiral fermion current is 
\beqn 
\nn
-\< \nabla_\m (\bar\psi^i iQ\delta_{ij} \g^\m L\psi^j )\> &=&
\frac{1}{24\pi^2} \e^{\m\n\r\s} {\rm tr}\Big[
Q^3 \pa_\m C_\n \pa_\r C_\s + \frac14 Q \pa_\m K_\n \pa_\r K_\s
+ Q^2 \pa_\m K_\n \pa_\r C_\s \\
&& \hspace{2.4cm}
- Q \Big( \pa_\m A_\n \pa_\r A_\s + \frac12 \pa_\m ( A_\n A_\r A_\s)
\Big)
\Big] \ ,
\lab{conanab1}
\eeqn
which contains the $U(1)^3$, mixed $U(1)-$K\"ahler
and mixed $G^2-U(1)$ anomalies. 
The anomaly of the last term of \reef{con} is
\beqn
\nn
\frac12 \xi \<\nabla_\m N^\mu \>
&=& -\frac{1}{24\pi^2}\frac12 \xi \e^{\m\n\r\s} \bigg[
- {\rm tr}(Q^2) \pa_\m C_\n \pa_\r C_\s
- {\rm tr}(Q) \pa_\m C_\n \pa_\r K_\s \\
&&\lab{conanab2}
\hspace{1cm}
+ \frac{1}{4}\Big(n_\l+3 - n_\psi\Big)\pa_\m K_\n \pa_\r K_\s
\\ \nn
&&
\hspace{1cm}
+ \Big( C_2(G) - \sum_r C(r)\Big) \Big( \pa_\m A_\n^a \pa_\r A^a_\s
  + \frac14 f^{abc} \pa_\m ( A^a_\n A^b_\r A^c_\s ) \Big)
\bigg]
\ .
\eeqn
In the $\pa K \pa K$ term, $n_\l={\rm dim}(\cg)$ is the total number of
gauginos, $n_\psi$ is the number of chiral fermions, and $3$ is the
gravitino contribution.

%%%%%%%%%%%%%%%%%%%%%%%%%%%%%%%%%%%%%%%%%%%%%%%%%%%%%%%%%%%%%%%%%

\subsection{Anomaly cancellation with local counter terms}
\label{s:sugraCTs}

As discussed in section \ref{s:lct}, non-gauge invariant local
counter terms can remove or restructure the
anomalies. We now apply the result of section \ref{s:lct} to the
anomaly conditions in the previous subsection. 

To start with consider the non-abelian consistency condition
\be\lab{connonab}
  -\< \nabla_\m (\bar\psi^i T^{a}_{ij} \g^\m L\psi^j )\>
  + \frac 12 \<\nabla_\m ( \bar\l^b f^{abc} \g^\m \l^c) \>
  = 0 \, ,
\ee
with the two contributions given by \reef{conan2a} and \reef{conan2b} above.
A counter term  
\be
  \label{CAA}
  \cl_\rom{CAA} = \frac{1}{12\pi^2} 
  \, \e^{\mu\nu\rho\sigma} C_\mu \tr\Big[ Q
  \Big( A_\nu\pa_ \rho A_\sigma 
  +\frac{3}{4} A_\nu A_ \rho A_\sigma \Big)\Big] 
\ee
will cancel the $G^2-U(1)$ mixed non-abelian anomaly and promote the
abelian mixed anomaly to covariant form. The mixed
$G^2-$K\"ahler anomaly is analogous, except for an overall
factor $C_2(G)-\sum_r C(r)$. The correct counter term is  
\be
  \label{KAA}
  \cl_\rom{KAA} = \frac{1}{24\pi^2} \Big( C_2(G)- \sum_r C(r) \Big)
  \e^{\mu\nu\rho\sigma} K_\mu
  \Big( A^a_\nu\pa_ \rho A^a_\sigma 
  +\frac{3}{8} f^{abc} A^a_\nu A^b_ \rho A^c_\sigma \Big) \, .
\ee
With these counter terms the non-abelian consistency condition becomes
\beqn\lab{conab}
0&=&- \theta^a \< \nabla_\m (\bar\psi^i T^{a}_{ij} \g^\m L\psi^j )\>
  + \frac 12 \theta^a \<\nabla_\m ( \bar\l^b f^{abc} \g^\m \l^c) \> 
  + \d_\theta \cl_{\rm CAA}+\d_\theta \cl_{\rm KAA} \non
  &=&
  \frac{i}{24\pi^2} \e^{\m\n\r\s} \theta^a\, \tr\,T^a 
  \Big[ \pa_\m A_\n \pa_\r A_\s + \frac12 \pa_\m ( A_\n A_\r A_\s )
  \Big] \, . 
\eeqn
Mixed terms have been removed and we are left with
the unavoidable non-abelian $G^3$ anomaly. Its cancellation
imposes the condition tr$[T^a\{T^b,T^c\}]=0$ on the matter spectrum. 

Let us now turn to the abelian consistency condition 
\beqn 
-\< \nabla_\m (\bar\psi^i iQ\delta_{ij} \g^\m L\psi^j )\> + 
\frac12 \xi \<\nabla_\m N^\mu \> ~=~ 0 \ . 
\eeqn 
The abelian gauge variations read 
\beqn 
\d_\L C_\mu = \pa_\mu \L\ , \quad 
\d_\L K_\mu = \xi \pa_\mu \L\ , 
\eeqn 
and the counter terms $\cl_\rom{CAA}$ and $\cl_\rom{KAA}$ restructure
the consistent mixed anomalies into gauge invariant form. 

The abelian consistency condition becomes  
\beqn
  \nonumber
   0&=&- \L \< \nabla_\m (\bar\psi^i iQ\delta_{ij} \g^\m L\psi^j )\> 
   +\frac12 \xi  \L \<\nabla_\m N^\mu \> 
   + \d_\L \cl_\rom{CAA} 
   +  \d_\L \cl_\rom{KAA}\\[2mm]\nn
   &=&
   \frac{1}{96\pi^2} \e^{\m\n\r\s} \L \Bigg[ \tr\, \Big[ \Big( Q +
       \frac12 \xi\Big) Q^2 \Big] C_{\m\n} C_{\r\s} 
+ {\rm tr}\Big[ \Big( Q +\frac12 \xi\Big) Q \Big] K_{\m\n} C_{\r\s}
\\[2mm]\nn
&&
\hspace{2.5cm}
+\frac14 \bigg( {\rm tr}\Big[ Q +\frac12 \xi \Big]
- \frac{1}{2}\xi \big( n_\l +
3\big) \bigg) K_{\m\n} K_{\r\s}
\\[2mm]
&&
\hspace{2.5cm}
- 3 \Big( {\rm tr}\Big[ \Big(Q+\frac12\xi\Big) T^aT^b \Big] + \frac12 \xi
C_2(G)\d^{ab} \Big) F^a_{\m\n} F^b_{\r\s}
\Bigg]
\, , 
\label{abelconX}
\eeqn
with $K_{\m\n} = \pa_\m K_\n - \pa_\n K_\m$. 

We must also consider the two counter terms
\beqn\lab{ctab}
\cl_\rom{CKK} = -\frac{1}{24\pi^2}\epsilon^{\m\n\r\s} C_\m K_\n \pa_\r K_\s \ , \quad\quad
\cl_\rom{KCC}=  -\frac{1}{24\pi^2}\epsilon^{\m\n\r\s} K_\m C_\n \pa_\r C_\s \ ,
\eeqn
which allow further cancellation of anomalies in the abelian
consistency condition.
Their gauge variations (after integration by parts) are
\beqn
\d_\L \cl_\rom{CKK}&=&  
-\frac{1}{96\pi^2}\epsilon^{\m\n\r\s}\L 
( \xi C_{\m\n} K_{\r\s} - K_{\m\n} K_{\r\s}) \ ,
\non
\d_\L \cl_\rom{KCC} &=& 
-\frac{1}{96\pi^2}\epsilon^{\m\n\r\s}\L ( C_{\m\n} K_{\r\s} - \xi C_{\m\n} C_{\r\s}) \ .
\eeqn
We add $a_\rom{CKK} \cl_\rom{CKK}+a_\rom{KCC} \cl_\rom{KCC}$ to the
Lagrangian and list below the independent terms from \reef{abelconX}
and \reef{conab}: 
\beqn
\nn
C \tilde{C}\, : && 0 =
{\rm tr}\Big[ \Big( Q +\frac12 \xi\Big)Q^2\Big] + \xi\, a_\rom{KCC} \ ,
\\[1mm]\nn
C \tilde{K}\, : && 0={\rm tr}\Big[ \Big( Q +\frac12 \xi\Big)
Q \Big] - \xi\, a_\rom{CKK} - a_\rom{KCC}  \ , 
\\[1mm]\nn
K \tilde{K}\, : && 0={\rm tr}\,Q  -
\frac12 \xi (n_\l + 3-n_\psi) + 4a_\rom{CKK}\ , 
\\[1mm]\nn
F^a \tilde{F}^b \, : && 0={\rm tr}\Big[ 
Q T^aT^b\Big]+\frac12 \xi \Big[ C_2(G) - \sum_r C(r)\Big]\d^{ab} \ ,
\\[2mm] 
G^3 \, : && 0={\rm tr} \Big[T^a\{T^b,T^c\} \Big]\ .
\label{noGScond}
\eeqn
Each of the conditions \reef{noGScond} must be satisfied separately.
Given the original field content of the model, these are the final and
physical conditions for the cancellation of gauge
anomalies. Gravitational anomalies will be considered next, and we
will add new fields required by the Green-Schwarz mechanism in section
\ref{s:GS}.  

There are additional consistency conditions from the gravitational
anomalies with and external gauge current and two energy-momentum
tensors in the triangle diagram. The resulting consistency condition
is
\beqn
  \nonumber
  0 &=& -\< \nabla_\m ( \bar\psi^i iQ \d_{ij} \g^\m L \psi^j \>_{\rm grav}
  +\frac{1}{2} \xi \< \nabla_\m N^\n \>_{\rm grav} \\[2mm]
  &=&
 -\frac1{768\pi^2} 
  \bigg[ 
    {\rm tr}(Q) - \frac{1}{2} \xi ( n_\l - 21 - n_\psi)
  \bigg]
  \e^{\m\n\r\s} R_{\m\n\x\t}R_{\r\s}{}^{\x\t}\ .
  \label{grav}
\eeqn

For models with $\xi=0$, one can choose $a_\rom{KCC}$ and $a_\rom{CKK}$
to satisfy the second and third conditions of \reef{noGScond}. The
remaining conditions reduce to the conventional anomaly cancellation
conditions of a gauge theory coupled to gravity, namely the
four traces $\tr\, Q$, $\tr\, Q^3$, $\tr\,  QT^aT^b$, and $\tr\,
T^a\{T^b,T^c\}$ must vanish.  
Note that without the counter term  contribution \reef{ctab}, the
second term in \reef{noGScond} would be positive definite and could
never be cancelled by adjustment of the matter field content.

We now consider models with non-vanishing \FI coupling. 
There are various cases of interest. 

The first case is just an abelian vector multiplet coupled to
  supergravity and no chiral multiplets \cite{Freedman:1976uk}. 
In this model $K_\mu = \xi C_\mu$; hence the counter terms
  \reef{ctab} vanish. The only gauge anomaly condition which remains
  is $0=\xi (n_\l+3) = 4 \xi$. The gravitational anomaly reduces to
  $0=\xi (n_\l-21) = -20 \xi$. Clearly the model is inconsistent for
  $\xi \ne 0$.

In general models, we now show that it is very unlikely that the
  consistency conditions can be satisfy for non-vanishing $\xi$.
  First we choose the
  counter term coefficients $a_\rom{KCC}$ and $a_\rom{CKK}$ to satisfy
  the first two conditions of \reef{noGScond} and substitute the value
  of $a_\rom{CKK}$ into the third condition. We then replace
  $n_\l-n_\psi$ by the value determined by \reef{grav}. The result is
\beqn
  0 &=& \tr\,\bigg[\Big(Q + \frac{1}{2}\xi \Big)\Big(Q + \xi \Big) Q \bigg]
        - 3 \xi^3 \, .
  \label{Qxicond}
\eeqn
Consistency now requires that we satisfy the $G^3$ and $F\tilde F$
conditions of \reef{noGScond}, and the conditions \reef{Qxicond} and
\reef{grav}. A solution would require that both conditions linear in
$\xi$ have a common solution which is then one of the roots of the
cubic condition \reef{Qxicond}. For given matter content this is
extremely unlikely.
This conclusion can be changed using a Green-Schwarz mechanism, as we
discuss in the next section.

%%%%%%%%%%%%%%%%%%%%%%%%%%%%%%%%%%%

%%%%%%%%%%%%%%%%%%%%%%%%%%%%%%%%%%%
\setcounter{equation}{0}
\section{Green-Schwarz anomaly cancellation}
\label{s:GS}

The Green-Schwarz mechanism for anomaly cancellation in four
dimensions is well known
\cite{Derendinger:1991hq,LopesCardoso:1991zt}. 
One adds a chiral multiplet with a gauged shift
symmetry. Decomposing the complex scalar $s$ of the chiral multiplet as
$s=\rho + i a$, the bosonic terms of the Green-Schwarz Lagrangian can
then be written as
\beqn \lab{action2}
\cl_{\rm GS} = 
-(\pa_\mu \rho)^2
-( \pa_\m a + c_{\rm GS} C_\m )^2
+ \frac1{96\pi^2}
 a\, \e^{\m\n\r\s}\pa_{\m} \O_{\n\r\s}\ . 
\eeqn
where the Chern-Simons form $\O_{\n\r\s}$ satisfies
\beqn\lab{Ome}
\hspace{-.5cm}
\e^{\m\n\r\s} \pa_{\m} \O_{\n\r\s} &=& 
\e^{\m\n\r\s} \Big[ b_{\rm CC} C_{\m\n}C_{\r\s} + b_{\rm CK}
  K_{\m\n}C_{\r\s} \nonumber \\
&& \hspace{1.5cm}
+ b_{\rm KK} K_{\m\n}K_{\r\s} + b_{\rm AA} F^a_{\m\n}F^a_{\r\s}
+ b_{\rm RR} R_{\m\n\eta\tau} R_{\r\s}^{~~\;\eta\tau}
\Big]\ .
\eeqn
In a model originating from string theory
\cite{Kaplunovsky:1994fg,Anastasopoulos:2006cz,Ibanez:1999pw, 
Klein:1999im,Scrucca:2000ns} (see also \cite{Louis:1996ya}), the
constants $b_{..}$ will 
be fixed, but we keep them arbitrary here to illustrate their role in
anomaly cancellation. 

The scalar $s$ is invariant under non-abelian gauge transformation, so
$\d_\theta \cl_\rom{GS}=0$. Under abelian gauge transformations,
\beqn
\label{varlittlea}
\d_\L a  = - c_{\rm GS} \L\ , \quad 
\d_\L \rho = 0 \ .
\eeqn
The first term in \reef{action2} is then gauge invariant, and it is
then the last term whose gauge variation modifies the previous
conditions \reef{noGScond} as follows: 
\beqn\lab{consfin}
C \tilde{C}\, : && 0=
{\rm tr}\Big[ \Big( Q +\frac12 \xi\Big)Q^2\Big] + \xi\, a_\rom{KCC}
- c_{\rm GS}\,  b_{\rm CC}  \ ,
\non
C\tilde{K}\, : && 0={\rm tr}\Big[ \Big( Q +\frac12 \xi\Big)
Q
\Big] - \xi \, a_\rom{CKK} -  a_\rom{KCC}  - c_{\rm GS}\,  b_{\rm CK}  \ , \non
K\tilde{K}\, : && 0={\rm tr}\,Q  -
\frac12 \xi (n_\l + 3-n_\psi) + 4a_\rom{CKK} - 4 c_{\rm GS}\,  b_{\rm
  KK}  \ , \non 
F^a \tilde{F}^b\, : && 0={\rm tr}\Big[ Q T^aT^b\Big]
 +\frac12 \xi\Big[ C_2(G) -\sum_r C(r)\Big]\d^{ab}
+ \frac{1}{3} c_{\rm GS}\,  b_{\rm AA} \d^{ab} \ ,
\non
G^3\, : && 0={\rm tr}\Big[T^a\{T^b,T^c\}\Big] \ , \nonumber \\
R \tilde{R}: && 0= 
   {\rm tr}(Q) - \frac{1}{2} \xi ( n_\l - 21 - n_\psi)
   + 8 c_{\rm GS}\,  b_{\rm RR} \, .
\eeqn
Here $n_\psi$ includes the contribution from the fermion
partner $\chi$ of the Green-Schwarz scalar $s$.
It is now evident that there is enough flexibility to cancel all
gauge anomalies, and the only condition that needs to be imposed on
the spectrum is ${\rm tr}[T^a\{T^b,T^c\}]=0$ for the irreducible
non-abelian anomaly. In fact, there is more flexibility than
needed. We can set $b_\rom{CK} = b_\rom{KK} = 0$ and thus
eliminate the composite connection completely from the Green-Schwarz
Lagrangian. The remaining parameters then suffice to cancel all but
the $G^3$ anomaly and allow an arbitrary value of the \FI coupling.

The Green-Schwarz mechanism has served well to cancel the anomalies,
but it has changed the physics of the model. To see this note that the
$a$-$C_\mu$ cross term can be removed 
by an appropriate gauge fixing condition. This leaves a mass term
\beqn 
-c_{\rm GS}^2 C_\m C^\m\ .
\eeqn 
Because of the gauged shift symmetry, the supersymmetric Lagrangian
also contains a gauge invariant mass term $c_\rom{GS} \bar{\lambda}_C
\chi$, where $\l_C$ is gaugino partner of $C_\mu$. This gives a
fermion mass equal to that of the gauge bosons.  

From the general form $D_\m s = \pa_\m s - X^{a s}A^a_\m$ we can
formally identify the Killing vector 
\beqn \lab{Kvec}
X^{s} = - i c_{\rm GS} \ . 
\eeqn 
The original scalars $z^\a$ and the Green-Schwarz scalar $s$ combine
in the gauge invariant K\"ahler potential
\beqn\lab{K2} 
K(s,\bar s,z,\bar z) = \d_{\a\bb}z^\a z^\bb + \frac12 (s+\bar
s)^2 \, .
\eeqn 
The new contribution from $s$ to the K\"ahler connection is gauge
invariant, so $\rom{Im}\, F^a = \xi$ is not changed, and the anomaly
analysis is unmodified.

From \reef{Fdef} with $F^a = i\xi$ we find the $U(1)$ D-term
\beqn 
D = i X^s K,_s + i X^\a K,_\a + \xi = \d_{\a\bb} z^\a Q z^\bb 
+2 c_{\rm GS} \r  + \xi \, . 
\eeqn 
The scalar potential of supergravity contains the term $\frac12 D^2$. 
This term is minimized at $D=0$, which can be achieved by adjusting
$\rho$. Thus the breaking of the $U(1)$ symmetry does not change the
vacuum energy.

%%%%%%%%%%%%%%%%%%%%%%%%%%%%%%%%%%%%%%%%%%%%%%%%%%%%%%%%%%%%%%%%%%%%
\setcounter{equation}{0}
\section{Generalizations}
\label{s:more}

\subsection{Non-gauge invariant K\"ahler potentials}
\lab{noninvK}

It is interesting to examine the effect of a gauge non-invariant
K\"ahler potential in the analysis of the gauge consistency
conditions. 
We consider as in \reef{flat1} the simplest model of
flat target space $\mathbb{C}^n$ with linearly realized gauge
symmetries. For simplicity, we exclude $U(1)$  factors and gauge
only a non-abelian simple subgroup of $SU(n)$.   
However, contrary to the gauge invariant K\"ahler potential
\reef{flat1} we take here the unconventional K\"ahler potential  
\beqn
  K(z,\bar{z}) = \d_{\alpha\bb} z^\a z^\bb + k(z) + \bar{k}(\bar{z})
  \, .
\eeqn
Here $k$ is a non-constant holomorphic function, whose gauge variation
\reef{Kgauge} generates  
\beqn
  F^a = X^{a\a} k_{,\a} \, .
\eeqn
It is somewhat artificial to break gauge symmetry by taking such
an unnatural K\"ahler potential. However, most of our analysis --- including
the Green-Schwarz cancellation mechanism --- applies to theories
on non-flat target spaces in which gauge symmetry breaking in the
K\"ahler potential cannot be avoided.

The gauge consistency conditions are again a special case of the
result of \cite{Freedman:2005up}. Since $\rom{Im}\, F^a \ne 0$ the
non-abelian consistency condition \reef{connonab} includes the
contribution from the divergence of the Noether current
and now reads
\beqn 
0 &=& 
-\< \nabla_\m (\bar\psi^i T^a_{ij} \g^\m L\psi^j )\>
+ \frac12 \< \nabla_\m ( \bar\l^b f^{abc} \g^\m \l^c) \>
+ \frac12 {\rm Im}\, F^a \< \nabla_\m N^\m \> \nn \\[2mm] \nn
&=&
\frac{1}{24\pi^2} \e^{\m\n\r\s} \Bigg[ 
\tr\, iT^a \Big\{ \pa_\m A_\n \pa_\r A_\s + \frac12 \pa_\m ( A_\n A_\r A_\s ) \Big\} \\[1mm]
&& \hspace{1.9cm}\nn
 - \widehat{\tr}\, T^a \Big\{ \pa_\m K_\n \pa_\r A_\s + \frac14  \pa_\m ( K_\n A_\r A_\s )
\Big\}  \\[1mm]
&& \hspace{1.9cm}\nn
- \frac{1}{2} \rom{Im}\,F^a\;
   \widehat{\tr}\,  \Big\{\pa_\m A_\n \pa_\r A_\s
  + \frac12 \pa_\m ( A_\n A_\r A_\s ) \Big\} \\[1mm]
&& \hspace{1.9cm}
  - \frac{1}{8} \rom{Im}\,F^a\; (n_\l+3-n_\psi)  \pa_\m K_\n \pa_\r K_\s
\Bigg] \, .
\label{ImFcon}
\eeqn
In the second equality we have inserted the expressions for the
anomalies \reef{conan2a}, \reef{conan2b}, and \reef{conanab2} with the
abelian $\xi$ replaced by $\rom{Im}\, F^a$. We have also dropped the
$U(1)$ contributions by setting $Q=0$.
For brevity, we have introduced the notation  
$\widehat{\tr} = \tr-\tr_\rom{adj}$, i.e.\ the trace over chiral
fermions minus the trace over gauginos.
The minus sign in $\widehat{\tr}$ arises in \reef{ImFcon} because the
K\"ahler connection couples to chiral fermions and gauginos with
opposite signs.  

Before discussing the gauge consistency conditions \reef{ImFcon} and
the effect of local counter terms, we comment on 
Wess-Zumino (WZ) consistency \reef{WZcond}. First recall that in the
analysis of the previous sections, $\rom{Im}\,F^a$ vanished so that
$K_\m$ was invariant under non-abelian gauge 
transformations. The first two lines of \reef{ImFcon} were then the only
contributions to the gauge consistency condition and each of them
independently 
satisfied the WZ consistency condition \reef{WZcond}. In the present
model, though, the K\"ahler connection does transform under non-abelian 
gauge transformations, 
$\d_\theta K_\m = \pa_\m (\rom{Im}\,F^a \theta^a)$,
and that gives an extra contribution to the variation of the anomaly
in the second line of \reef{ImFcon}; by itself the second line of
\reef{ImFcon} is no longer WZ consistent. 

It turns out that WZ consistency is saved by contributions
from  $\frac{1}{2}\rom{Im}\,F^a \< \nabla_\m N^\m \>$.
The $\partial K \partial K$
contribution in the last line of \reef{ImFcon} satisfies WZ
consistency because of the non-abelian transformation of
$\rom{Im}\,F^a$. The gauge variation of the third line in
\reef{ImFcon} has two contributions: the variation of 
$\rom{Im}\,F^a$ yields the term required by WZ consistency, but the
variation of $\widehat{\tr}\,[dAdA+(1/2)d(A^3)]$
gives an extra term. Conveniently, 
that extra term precisely cancels the unwanted term from the variation of
$K_\m$ in the second line of \reef{ImFcon}. Thus the full expression
\reef{ImFcon} does indeed satisfy the WZ consistency condition
\reef{WZcond}. 

Returning to the analysis of the gauge consistency condition
\reef{ImFcon} we note that since $K_\mu$ transforms as an abelian
connection under non-abelian gauge variations, the local counter term 
$\cl_\rom{KAA}$ given in \reef{KAA} removes the anomaly in the second
line of \reef{ImFcon} and it simultaneously converts the third line to
covariant form. Including $\cl_\rom{KAA}$ in the Lagrangian, the
physically relevant form of the gauge consistency condition becomes
\beqn
0 &=& 
-\< \nabla_\m (\bar\psi^i T^a_{ij} \g^\m L\psi^j )\>
+ \frac12 \< \nabla_\m ( \bar\l^b f^{abc} \g^\m \l^c) \>
+ \frac12 {\rm Im}\, F^a \< \nabla_\m N^\m \> 
+ \d^a_\theta \cl_\rom{KAA} \nn \\[2mm] 
\label{ImFconCT}
&=&
\e^{\m\n\r\s} \Bigg[ 
\frac{1}{24\pi^2} \tr\, iT^a \Big\{ \pa_\m A_\n \pa_\r A_\s 
+ \frac12 \pa_\m ( A_\n A_\r A_\s ) \Big\} 
\\[1mm]
&& \hspace{1.9cm}\nn
- \frac{1}{32\pi^2}\, \frac{1}{2} \rom{Im}\,F^a\;
   \widehat{\tr}\,  F_{\m\n} F_{\r\s}  -
   \frac{1}{96\pi^2}\,\frac{1}{2} \rom{Im}\,F^a\;
   \frac{1}{4}(n_\l+3-n_\psi)   K_{\m\n} K_{\r\s} 
\Bigg] \, .
\eeqn
For non-vanishing $\rom{Im}\,F^a$, consistency requires, besides the
usual $G^3$ anomaly condition, that $C_2(G)=\sum_r C(r)$ and
$n_\l+3-n_\psi=0$. In addition, cancellation of the gravitational anomaly
requires $n_\l-21-n_\psi=0$. It is clear that these conditions
cannot be simultaneously be satisfied, and so the models are
inconsistent. An example is the 
(non-supersymmetric) gaugino model considered as a toy example in
\cite{Freedman:2005up}. The model has no chiral multiplets and no
gravitino, and the gauge anomaly $K\tilde{K}$ renders the model
inconsistent. 

In the previous section, we successfully applied the Green-Schwarz
mechanism to remove anomalies. The fact that $K_\m$ transforms as a
$U(1)$ connection under non-abelian gauge transformations, 
suggests that a Green-Schwarz mechanism can remove
covariant anomalies proportional to $\rom{Im}\, F^a$ from the
non-abelian conservation law.

A Green-Schwarz mechanism with a Chern-Simons term 
$a\, \e^{\m\n\r\s} \pa_\m \Omega_{\n\r\s}$ 
can be used to cancel the mixed non-abelian anomalies
in \reef{ImFconCT} provided that the axion transforms under
non-abelian gauge transformations as
$\d_\theta a = - k_\rom{GS}\, \rom{Im}\, F^a \theta^a$
for some constant $k_\rom{GS}$, see \cite{Derendinger:1991hq}. Holomorphic behavior of the 
Green-Schwarz complex scalar $s=\rho + ia$ requires that it transforms
non-trivially as 
\be
  \label{ds}
  \d_\theta s = - k_\rom{GS}\, F^a \theta^a \, .
\ee
The gauge invariant supersymmetric kinetic term for $s$ is obtained
from the superfield K\"ahler potential
\be
  \label{supernew}
  \frac{1}{2}(S+\bar{S}+k_\rom{GS}\, K^{(0)})^2 \, ,
\ee
where $S$ is the chiral superfield whose lowest component is $s$ and
$K^{(0)}=K^{(0)}(Z,\bar{Z},V)$ is the standard K\"ahler potential for
the chiral superfields $Z$, involving the real superfield $V = V^a
T^a$ of the vector multiplet. 
The full K\"ahler potential is now
\be
  \label{KnewGS}
  K(z,\bar{z},s,\bar{s}) = K^{(0)}(z,\bar{z}) 
  + \frac{1}{2} \Big( s+\bar{s} + k_\rom{GS} K^{(0)}(z,\bar{z}) \Big)^2 \, .
\ee
We label the original K\"ahler potential and
metric with superscripts $^{(0)}$, 
i.e.~$G^{(0)}_{\a \bb} = K^{(0)}_{,\a\bb}$.   
It follows from \reef{KnewGS} that the scalar kinetic terms are
\bea
  \nn
  &&
  \hspace{-2cm} 
  -G_{\a \bb}D_\mu z^\a  D^\mu z^\bb
  -G_{\a \bar{s}} D_\mu z^\a D^\mu \bar{s}
  -G_{s \bb} D_\mu s D^\mu z^\bb
  -G_{s \bar{s}} D_\mu s D^\mu \bar{s} \\[2mm]
 \nn
 &=&
  -G^{(0)}_{\a \bb}D_\mu z^\a  D^\mu z^\bb
  - k_\rom{GS} \big(s + \bar{s} + k_\rom{GS} K^{(0)} \big) 
G^{(0)}_{\a\bb} D_\mu z^\a  D^\mu z^\bb  \\[2mm]
 &&  
  - (D_\mu s + k_\rom{GS} K^{(0)}_{,\a} \, D_\mu z^\a)
    (D^\mu \bar{s} + k_\rom{GS} K^{(0)}_{,\bb} \, D^\mu z^\bb) \, ,
 \label{Kkin}
\eea 
where we have used that \reef{ds} implies 
${X^a}^s=- k_\rom{GS} F^a$ for the holomorphic Killing vector, so that
$D_\mu s = \pa_\m s + k_\rom{GS}\, F^a\, A^a_\mu$.
The first term in \reef{Kkin} is just the standard $z$-$\bar{z}$
kinetic term, and the two other terms come from the 
Green-Schwarz Lagrangian. 
Using the identity
\be
  \frac{1}{2} \pa_\m K^{(0)} + i K^{(0)}_\m = 
  K^{(0)}_{,\a} \pa_\m z^\a 
  + A^a_\m F^a - X^{a\a} K^{(0)}_{,\a} A_\mu^a
  \, ,
\ee
where $K^{(0)}_\m$ is the original K\"ahler connection,
we rewrite the last term of \reef{Kkin}. Then the Green-Schwarz
Lagrangian for the chiral scalars takes the form
\bea
  \nn
  \cl_\rom{GS} &=& -\Big(\pa_\m a + k_\rom{GS} K^{(0)}_\m\Big)^2 
 -\bigg [\pa_\m \Big(\rho + \frac{1}{2}k_\rom{GS} K^{(0)} \Big) \bigg]^2 \\
  &&
 - k_\rom{GS} \Big(2\rho + k_\rom{GS} K^{(0)} \Big)
   G^{(0)}_{\a\bb} D_\mu z^\a  D^\mu z^\bb 
   + \frac{1}{96\pi^2}\, a\, \e^{\m\n\r\s} \pa_\m \Omega_{\n\r\s} \, .~ 
\eea
Note that the Green-Schwarz Lagrangian includes a correction to the
$z$-$\bar{z}$ kinetic term. 

The Green-Schwarz scalars and the new term in the 
K\"ahler potential \reef{KnewGS} contribute to the K\"ahler
connection, giving
\be
  \label{GSKconn}
  K_\mu = K^{(0)}_\mu + \Big(2\rho + k_\rom{GS} K^{(0)}\Big)
  \Big(\pa_\mu a +  k_\rom{GS} K^{(0)}_\mu\Big) \, .
\ee
Let the Chern-Simons form $\O_{\n\r\s}$ satisfy
\beqn \label{thisOme}
\hspace{-.5cm}
\e^{\m\n\r\s} \pa_{\m} \O_{\n\r\s} &=& 
\e^{\m\n\r\s} \Big[ 
b_{\rm KK} K_{\m\n}K_{\r\s} + b_{\rm AA} F^a_{\m\n}F^a_{\r\s}
+ b_{\rm RR} R_{\m\n\eta\tau} R_{\r\s}^{~~\;\eta\tau}
\Big]\ ,
\eeqn
with $K_{\m\nu}$ the field strength of the corrected K\"ahler
connection \reef{GSKconn}. 
Since the gauge invariant correction to the K\"ahler
potential \reef{KnewGS} does not change the value of 
$\rom{Im}\, F^a$, the constants $b_{..}$ in
\reef{thisOme} can be chosen to cancel the mixed anomalies in
\reef{ImFconCT} as well as the gravitational anomaly 
proportional to $\rom{Im}\, F^a$. This leaves only the usual $G^3$
anomaly. We conclude that the models
considered in this section can be consistent only when the
Green-Schwarz mechanism with the composite connection is included.  

As for the standard  Green-Schwarz mechanism, there are corrections to
the D-term potential. We find
\be
  D^a = {D^{(0)}}^a \left[ 1+ k_\rom{GS} (2\rho + k_\rom{GS} K^{(0)})\right]
  \, ,
\ee
where ${D^{(0)}}^a$ is the D-term before the corrections from the
Green-Schwarz mechanism. Again the D-term conditions can be solved by
adjusting $\rho$ to make $D^a=0$.

%%%%%%%%%%%%%%%%%%%%%%%%%%%%%%%%%%%%%%%%%%%%%%%%%%%%%%%%

\subsection{Supersymmetrization} 
\lab{appsusy}

So far our analysis has only included the bosonic terms of the
anomalies. Since we started with a supersymmetric theory it is natural
to consider supersymmetrized forms of local counter terms and the
Green-Schwarz mechanism. 
Supersymmetric versions of the Green-Schwarz Lagrangian are known
\cite{Derendinger:1991hq,LopesCardoso:1991zt}, so
here we focus on the counter terms.  

The consistent anomaly is not gauge invariant, so we cannot work in
Wess-Zumino gauge and must resort to superfields. The superfield
version of the covariant anomaly is straightforward, 
$\mathcal{A}_\rom{cov} \propto \int d^4x\, d^2\th \tr\, i \L W^\a W_\a
+ h.c.$, 
where $W^\a$ denotes the non-abelian superfield vector field strength 
and $\L$ is a chiral superfield.\footnote{In this section, we focus on
  global supersymmetry. We use
  spinor indices $\a, \dot\a$ for the components of Weyl fermions. 
We use standard superspace conventions, for relevant details see
\cite{Marinkovic:1990ny,Ohshima:1999jg}.}   
Supersymmetric expressions for the difference between the consistent
and covariant anomaly for a simple gauge group are complicated
\cite{Marinkovic:1990ny,Ohshima:1999jg,Gates:2000dq,Gates:2000gu}.
Some simplication occurs for the mixed $U(1)-G^2$ abelian anomaly
which can be obtained from \cite{Marinkovic:1990ny,Ohshima:1999jg} and
written as 
\beqn
  &&\mathcal{A}^Q_\rom{mixed~con}-
  \mathcal{A}^Q_\rom{mixed~cov}
  =
 \frac{1}{64\pi^2}
 \int d^4x\, d^4\th 
  \int_0^1 dg \, 
  \delta_\Lambda C\, 
   \tr \,
  \Big[ 
   Q\, 
   X_g(A)
    \Big] \, ,\nonumber \\
   \label{susyAnom} 
   \\ \nonumber
  &&X_g(A) =
   \bigg(
      [\mathcal{D}^\a A, W_\a(A)]
      + [\bar{D}_{\dot\a} A, \bar{\mathcal{W}}^{\dot\a}(A)]
      + \{ A, \mathcal{D}^\a W_\a(A)\}
    \bigg)_{A\rightarrow gA} \, .
\eeqn

As observed in \reef{abelvarLct}, a desired property is that the abelian variation of the counter term restructures the mixed abelian consistent anomaly to covariant form. From \reef{susyAnom} we can directly read off that the counter term
\beqn 
\label{susyCT}
  \mathcal{L}_\rom{Sct}(A,C)
&=&
  -\frac{1}{64\pi^2}
  \int d^4\th \int_{0}^{1} dg \,
  C \,
  \tr \, \Big[
  Q  \, X_g(A) \Big]
\ . 
\eeqn
has exactly this property.
Note that this fixes the counter term only up to terms that are
invariant under abelian transformations.  
The non-abelian variation of the counter term \reef{susyCT} should
cancel the supersymmetrized version of the mixed $U(1)-G^2$ non-abelian
anomaly $\mathcal{A}^a_\rom{mixed~con}$.
Due to the complicated structure of the non-abelian variation
$\d_\Theta A$, we have only confirmed the cancellation of
$\mathcal{A}^a_\rom{mixed~con}$ at leading 
order.\footnote{The form of the consistent anomaly presented in
\cite{Gates:2000dq,Gates:2000gu} may be more useful for
this purpose since the expressions there involve only $e^A$ and
the complications of the non-abelian gauge variations $\delta_\Theta A$ do not 
arise.}
However, in component form, the
supersymmetric counter term \reef{susyCT} correctly reproduces the
bosonic counter term \reef{genSct}.

There is another approach to the mixed consistent anomaly involving
the descent equations.\footnote{We thank Massimo Bianchi and Emilian
  Dudas for drawing our attention to this point, see
  \cite{Anastasopoulos:2006cz}.}  
This leads to a different form of the mixed consistent anomalies,
and suggests a counter term involving the non-abelian 
Chern-Simons three-form whose superfield expression can be found in
\cite{Cecotti:1987nw,Derendinger:1991hq}. The resulting superfield
counter term is similar to \reef{susyCT}.

We have discussed some issues associated with
a superfield formulation of counter terms with the desired properties. 
Addition study is needed to recast the consistency conditions and the
full structure of the local counter terms in manifestly supersymmetric
form.

%%%%%%%%%%%%%%%%%%%%%%%%%%%%%%%%%%%%%%%%%%%%%%%%%

\setcounter{equation}{0}
\section{Conclusions}
\label{s:concl}

We have clarified the consistency conditions that follow from the current 
conservation law \reef{con2} in supergravity \cite{Freedman:2005up} for 
flat sigma-model target space and linearly realized gauge
symmetries. The analysis shows that 
anomalies arising from the non-invariance of the composite K\"ahler
connection under gauge transformations complicate the anomaly
cancellation conditions. 

Starting from the consistent anomaly
\cite{Bardeen:1969md,Gross:1972pv} and including all finite local
counter terms we reduce the consistency conditions to a set of
physically relevant conditions. For vanishing \FI couplings the
conditions simplify to the standard anomaly cancellation conditions
well-known from the Standard Model or the MSSM. However, a
non-vanishing \FI coupling $\xi$ gives more involved consistency
conditions. The usual $G^3$ condition $\rom{Tr}\,T^a\{ T^b, T^c \} =0$
must hold, but the \FI coupling modifies the cancellation of the
gravitational and 
the mixed $U(1)-G^2$ anomalies. 
Some anomalies may be removed by including finite local
counter terms in the action. A consistent model then requires: 
\beqn
\nonumber
G^3 \, : && 0={\rm tr} \Big[T^a\{T^b,T^c\} \Big]\ , \\[2mm] 
\nonumber
  F^a \tilde{F}^b \, : && 0={\rm tr}\Big[ 
Q T^aT^b\Big]+\frac12 \xi \Big[ C_2(G) - \sum_r C(r)\Big]\d^{ab} \ ,
\\[2mm]
\nonumber 
R \tilde{R} \, : 
&& 0={\rm tr}(Q) - \frac{1}{2} \xi ( n_\l - 21 - n_\psi)\, , \\[2mm] 
\rom{abelian} \, : 
&& 0 = \tr\,\bigg[\Big(Q + \frac{1}{2}\xi \Big)\Big(Q + \xi \Big) Q
  \bigg] - 3 \xi^3 \, .
\eeqn
Solving the full set of
conditions to find a consistent model looks unlikely, nonetheless it
would be curious to see if there actually are consistent models. It
is clear that the \FI coupling in such a model cannot be treated as
an arbitrary parameter.  

At the cost of including extra degrees of freedom, the
Green-Schwarz mechanism provides enough flexibility to cancel
anomalies for arbitrary values of the \FI couplings.  

An immediate consequence of adding the Green-Schwarz Lagrangian is a
mass term for the abelian gauge boson and 
a modification of the D-term potential from the Green-Schwarz
scalar. In other words, in the 
presence of a \FI coupling the abelian gauge boson always gains a mass
irrespective of the vacuum structure. Furthermore, the D-term is
corrected by the contribution from the Green-Schwarz scalar $s$. It
then reads
\beqn 
\label{Dpotential}
\frac12 D^2 = \frac12 \Big( \sum_i q_i |\phi_i|^2 + \xi +
c_\rom{GS} K_{,s} \Big)^2 \ , 
\eeqn 
for linearly transforming matter fields $\phi_i$ with $U(1)$ charges
$q_i$. The $c_\rom{GS}$-term enters through the Green-Schwarz Lagrangian.
Formally, this correction is a one-loop effect, and there may be more
corrections at 
the same order in perturbation theory that affect the scalar potential.
This means there are effectively no field-independent \FI
couplings.\footnote{The possibility to ``integrate out'' the new
scalar by setting it to a constant background value in the D-term
seems only feasable if supersymmetry gets broken 
in the process, see \cite{Binetruy:2004hh}.}

As a generalization we also study consistency conditions
for models with non-invariant K\"ahler potentials. 
Again the models are assumed to have flat target space and linearly
realized gauge symmetries. We focus on models with simple
gauge groups, leaving the inclusion of $U(1)$-factors as a possible
generalization. 
While local counter terms remove mixed K\"ahler$-G^2$
anomalies from the non-abelian consistency condition, the
conditions for cancelling $K\tilde{K}$
and gravitational anomalies cannot simultaneously be satisfied, and so
the models are inconsistent as they stand.

The Green-Schwarz mechanism based on the composite connection and the
K\"ahler potential can remove this type of non-abelian anomalies.
Since the K\"ahler connection transforms as
an abelian connection under non-abelian gauge variations, it is
possible to arrange a Green-Schwarz Chern-Simons term to cancel the
anomalies in the non-abelian current conservation law. 
The simple models with non-invariant K\"ahler
potentials can be consistent only when this Green-Schwarz
mechanism is included to cancel the anomalies. As in
\reef{Dpotential}, there are corrections to the D-term potential.

%%%%%%%%%%%%%%%%%%%%%%%%%%%%%%%%%

\newpage
\begin{center}
{\bf Acknowledgements}
\end{center}
\vspace{-.5cm}

We would like to thank Luis \'Alvarez-Gaum\'e, Massimo Bianchi,
Emilian Dudas, Gia Dvali, Michael Haack, Roman Jackiw, and Angel
Uranga.  
The research of D.~Z.~F.\ is supported
by the NSF grant PHY-00-96515. Further support comes from funds provided by
the U.S. Department of Energy (D.O.E.) under cooperative research
agreement $\#$DF-FC02-94ER40818. H.E.~was supported by a Pappalardo
Fellowship in Physics at MIT and by the US Department of Energy
through cooperative research agreement DF-FC02-94ER40818.

%%%%%%%%%%%%%%%%%%%%%%%%%%%%%%%%%%%%%%%%%%%%%%%%%%%%%%%%%%%%%%%%%%%%%%%%

\end{document}